\documentclass{IEEEtran}
\usepackage{cite}
\usepackage{amsmath,amssymb,amsfonts}
\usepackage{algorithmic}
\usepackage{graphicx}
\usepackage{textcomp}
\usepackage[compress,capitalize]{cleveref} 
\crefname{figure}{Fig.}{Figs.}

\def\BibTeX{{\rm B\kern-.05em{\sc i\kern-.025em b}\kern-.08em
    T\kern-.1667em\lower.7ex\hbox{E}\kern-.125emX}}
\begin{document}
\title{Continuous PPG-Based Blood Pressure Monitoring Using Multi-Linear Regression}
\author{Serj Haddad, Assim Boukhayma, and Antonino Caizzone
\thanks{This paper was submitted for review on November 19, 2020. This work was supported by Senbiosys, Neuchatel 2002, Switzerland (e-mail: serj.haddad@senbiosys.com). }}

\maketitle

\begin{abstract}
In this work, we present a photoplethysmography-based blood pressure monitoring algorithm (PPG-BPM) that solely requires a photoplethysmography (PPG) signal. The technology is based on pulse wave analysis (PWA) of PPG signals retrieved from different body locations to continuously estimate the systolic blood pressure (SBP) and the diastolic blood pressure (DBP). The proposed algorithm extracts morphological features from the PPG signal and maps them to SBP and DBP values using a multiple linear regression (MLR) model. The performance of the algorithm is evaluated on the publicly available Multiparameter Intelligent Monitoring in Intensive Care (MIMIC I) database. We utilize 28 data-sets (records) from the MIMIC I database that contain both PPG and brachial arterial blood pressure (ABP) signals. The collected PPG and ABP signals are synchronized and divided into intervals of 30 seconds, called epochs. In total, we utilize  $47153$ \textit{clean} 30-second epochs for the performance analysis. Out of the 28 data-sets, we use only 2 data-sets (records $041$ and $427$ in the MIMIC I) with a total of $2677$ \textit{clean} 30-second epochs to build the MLR model of the algorithm. For the SBP, a mean difference ($\pm$ standard deviation) of $0.0 \pm 8.01$ mmHg and a mean absolute error (MAE) of $6.10$ mmHg between the arterial line and the PPG-based values are achieved, with a Pearson correlation coefficient $r = 0.90$, $p<.001$.  For the DBP, a mean difference of $0.0 \pm 6.22$ mmHg and an MAE of $4.65$ mmHg between the arterial line and the PPG-based values are achieved, with a Pearson correlation coefficient $r = 0.85$, $p<.001$. We also use a binary classifier for the BP values with the positives indicating SBP $\geq 130$ mmHg and/or DBP $\geq 80$ mmHg and the negatives indicating otherwise. The classifier results generated by the PPG-based SBP and DBP estimates achieve a sensitivity and a specificity of $79.11\%$ and $92.37\%$, respectively.
\end{abstract}


\section{Introduction}
%%%%%%%%%%%%%%%%%%%%%%%%%%%%%%%%%%%%%%%%%%%%%%%%%%%%%%%%%%%%%%%%%%%%
\label{sec:introduction}
This work presents a detailed description of how a photoplethysmography (PPG) signal can be used to estimate the blood pressure (BP). In what follows, we discuss: \ref{subsec:hypertension}) the importance of accurate continuous blood pressure monitoring, \ref{subsec:direct_methods}) the direct methods used to measure blood pressure, \ref{subsec:ppg_methods}) the PPG based methods used to estimate blood pressure, \ref{subsec:objectives}) the main objectives of this paper and \ref{subsec:paper_organization}) how it is organized.

\subsection{Hypertension}\label{subsec:hypertension}
Hypertension is a long-term medical condition in which the blood pressure in the arteries is persistently elevated. It is the main risk factor for cardiovascular diseases and a major cause of premature death worldwide. According to the World Health Organization (WHO), the number of people with hypertension around the world is estimated to be 1.13 billion. More than two-thirds of the population with hypertension live in average and low income countries. Unfortunately, only less than $20\%$ of hypertensive people have the problem under control. According to WHO, one of the global targets set for noncommunicable diseases is the reduction in the prevalence of hypertension by $25\%$ (baseline 2010) by the year 2025 \cite{WHO}.

Based on the key facts mentioned, the potential solution to fight hypertension must address the following issue: hypertension is largely asymptomatic, hence the name - silent killer, which requires accurate, long-term, and continuous monitoring for a large number of populations. This means that the proposed solution should be 1) inexpensive to make it available to the widest possible population, 2) unobtrusive and comfortable to enable continuous long-term monitoring during regular daily activities, and most importantly, 3) \textit{sufficiently} accurate compared to the clinically accepted BP monitoring devices.

\subsection{Direct Methods}\label{subsec:direct_methods}
The arterial line is an invasive method and the gold-standard to measure BP values continuously and accurately, yet it is not suitable for daily ambulatory use. On the other hand, cuff-based, auscultatory and oscillometric BP monitoring, which is the gold non-invasive method, is intermittent and occludes the blood flow in the arteries resulting in uncomfortable and unpleasant feelings \cite{Park2001}. For ambulatory studies, frequent cuff inflation and blood flow occlusion become painful and prevent a proper monitoring especially during sleep \cite{Davies1994}. 

\subsection{PPG-based Methods}\label{subsec:ppg_methods}
Photoplethysmography was proposed as a promising low-cost, non-invasive, and non-occlusive technique for continuous BP estimation \cite{Teng2003,Choudhurry2014}. PPG involves using a light-emitting diode (LED) to illuminate the skin  and measuring the intensity of either the transmitted or the reflected light to a photo-diode. This optical solution detects the volumetric changes in the blood flow. It is already available in many wearable devices and smart phones. Many studies confirm the accuracy of PPG-based solutions to monitor heart rate (HR) and heart rate variability (HRV) \cite{AccurateHRVAcharya,Serj2020}, stress levels \cite{StressSztajzel,StressThayer}, irregularity in heart beats \cite{EctopicSerj,EctopicTarniceriu}, sleep quality \cite{SleepRenevey2,SleepMyllymaki}, etc. For BP monitoring, PPG-based methods rely on either pulse wave velocity (PWV) techniques or pulse wave analysis (PWA) methods. 
\subsubsection{pulse wave velocity}
Arterial stiffness directly affects the blood pressure. The stiffening of the arteries increases the velocity of the pressure wave propagation through the arterial tree and results in an increased blood pressure. PWV techniques aim at estimating the blood pressure by computing the propagation velocity. Some of these techniques estimate the pulse transit time (PTT) \cite{PTT2013,PTT2014,PEP2015}, which is the time delay between proximal and distal arterial waveforms and is inversely proportional to the PWV. This can be achieved using two PPG sensors in two different body locations. Other techniques, known as pulse arrival time (PAT) techniques \cite{PAT2009, PAT2019}, replace the transit time estimation by the interval estimation from the electrocardiogram (ECG) R-peak to the pulse arrival at the peripherals. Different characteristic points on the PPG pulse waveform can be considered as the pulse arrival location \cite{Sola2009}. Since the time interval starts from the R-peak of the ECG, the PAT includes both the PTT and the cardiac pre-ejection period (PEP) \cite{PEP2015}. Despite these differences, the main goal of both techniques is to accurately compute the PWV since it is strongly correlated with the blood pressure. 

One of the main challenges for the PWV techniques is that the peripheral arterioles or the terminal arteries lack the elasticity that the central arteries have. For this reason, the PWV techniques fail to maintain the good performance they deliver for the central arteries when they are deployed at the body peripherals. Moreover, PPG-based BP monitoring solutions that rely on the PWV techniques require the use of auxiliary sensors, which limits its practicality and usability. 
\subsubsection{pulse wave analysis}
Contrary to the PPG-based PWV techniques, the PPG-based pulse wave analysis (PWA) relies solely on PPG signals \cite{PWA1, PWA2}. PWA techniques  involve a morphological analysis of the PPG pulse waveform to extract features that can be used to estimate the blood pressure. Several time-related and amplitude-related features are proposed in the state-of-the-art \cite{PWAFeaturesAtomi2017, PWAFeaturesShimazaki2018}. The extracted features are then mapped to blood pressure values using different techniques; such as, multiple linear regression (MLR) \cite{PWA_MLR_RF, PWA_MLR}, artificial neural networks (ANN) \cite{PWA_ANN1, PWA_ANN2}, support vector machine (SVM) \cite{PWA_SVM}, random forests (RF) \cite{PWA_MLR_RF}, etc. It is also worth mentioning that with the increasing interest in deep learning techniques, recent works have suggested PWA techniques using raw PPG signals as input to deep neural networks (DNN), without the need for explicit feature extraction \cite{PWA_DNN}.

\subsection{Objectives}\label{subsec:objectives}
The three main challenges of the blood pressure estimation with the PWA approach are: 1) the identification of the features, 2) the accurate extraction of the features, and 3) the mapping of the extracted features to blood pressure values. The PPG features used in PWA techniques are numerous and are widely studied in the literature. However, it is essential to identify the features that have an analogous significance/relevance over a wide spectrum of data-sets. Identifying features that have uniform correlation with blood pressure values over diverse data-sets ensures the generalization of the suggested PWA model. The feature extraction is also challenging for certain PPG morphology variations, especially those corresponding to the elderly. Finally, the most challenging aspect of the PWA approach lies in the feature mapping. Any regression model used by the PWA technique requires training data-sets for proper tuning of its parameters. Choosing the data-sets is very significant to ensure that the model generalizes well and maintains a comparable performance throughout varying data-sets.

The objectives of this work are to give effective insights into proper processing of publicly available data-sets, rigorous construction of regression models, and accurate performance evaluation that truthfully reflects the reality of the proposed scheme. To achieve our objectives, we elaborate in this work the following aspects. 
\subsubsection{data used to build the algorithm}
It is important to identify the data that contributed to the algorithm design. This helps identify the data-sets that are capable of providing sufficient information regarding different PPG morphology variations and different feature combinations. As such, we ensure that the algorithm maintains a good performance with a wide range of data-sets. Furthermore, using a small proportion of the available data-sets to construct the regression model of the algorithm alone does not indicate the robustness of the algorithm. It is important to know how the training data are distributed over different records. 

In this work, we consider the Multiparameter Intelligent Monitoring in Intensive Care (MIMIC I) database \cite{MIMICI}. The data from MIMIC I used to construct the MLR model of the PPG-BPM technology constitute only $5.68\%$ of the total MIMIC I data studied in this work. Moreover, this training data belong to only 2 records (records $041$ and $427$) out of the 28 records considered in this work. 
\subsubsection{good performance over different subjects/records}
The objective is to demonstrate the good performance of the proposed algorithm over data-sets belonging to different subjects. In \cref{Table:Subjects}, we illustrate the per subject performance of the algorithm in terms of the standard deviation of the error in the systolic blood pressure (SBP) and the diastolic blood pressure (DBP). The standard deviations of the error in SBP and DBP are $\sim 8$ mmHg and $\sim 6$ mmHg, respectively, for most of the subjects. This performance is similar to the overall performance of the algorithm (refer to \cref{Table:MIMIC}, the standard deviation of the error in SBP is $8.08$ mmHg and that of DBP is  $6.22$ mmHg).
\subsubsection{good performance over BP groups}
The objective is to demonstrate the good performance of the PPG-BPM algorithm over different BP groups. For this, we divide the arterial blood pressure (ABP) recordings into two groups: \textit{hypertensive} blood pressure (SBP $\geq 130$ mmHg and/or DBP $\geq 80$) and \textit{non-hypertensive} blood pressure. We justify this classification in \cref{subsec:ClassResults}. As presented in \cref{Table:Accuracy}, the algorithm achieves a classification accuracy of $88.14\%$. Moreover, as shown in \cref{Table:Classification}, the algorithm achieves a good performance for both the \textit{hypertensive} and the \textit{non-hypertensive} groups.
\subsubsection{good performance over different databases}
We further extend our performance analysis to cuff-based recordings. For this purpose, we utilize the University of Queensland Vital Signs data-set \cite{QUD}. In \cref{Table:QUD}, we illustrate how our algorithm maintains a good performance compared to cuff-based BP measurements.

\subsection{Paper Organization}\label{subsec:paper_organization}
\cref{sec:methods} presents the methods used in this study. This includes information about the participants of the study, the statistics of the acquired data, and the methods utilized to process the collected data. \cref{sec:algorithm} presents the different building blocks of the proposed blood pressure monitoring algorithm. In \cref{sec:results}, we evaluate the accuracy of the PPG-BPM technology to continuously monitor blood pressure compared to the invasive intra-arterial blood pressure values. \cref{sec:discussions} includes discussions about the results of the paper, the performance compared to the cuff oscillometric measurements, and the calibration process. Finally, \cref{sec:conclusion} summarizes our conclusions.

\section{Signal Processing \& Data Analysis}\label{sec:methods}
%%%%%%%%%%%%%%%%%%%%%%%%%%%%%%%%%%%%%%%%%%%%%%%%%%%%%%%%%%%%%%%%%%%%
In this section, we describe the methods used to process the data from the MIMIC I database: 1) the pre-processing of the PPG signal before running the algorithm and 2) the generation of the reference SBP and DBP values from the arterial line signal.

\subsection{MIMIC I Database}
In this work, we consider 28 records, listed in \cref{Table:Subjects}, of the 72 records available in the MIMIC I database. We consider the records that do not have missing, abnormal, or noisy signals. Moreover, we discard the records for which the proposed algorithm generates less than 15 minutes of BP estimates. We elaborate more on how the algorithm discards noisy PPG signals in \cref{sec:algorithm}.  
\subsection{Data Segmentation}\label{subsec:segmentation}
\begin{figure}[t]
\centering
\includegraphics[width=1\linewidth]{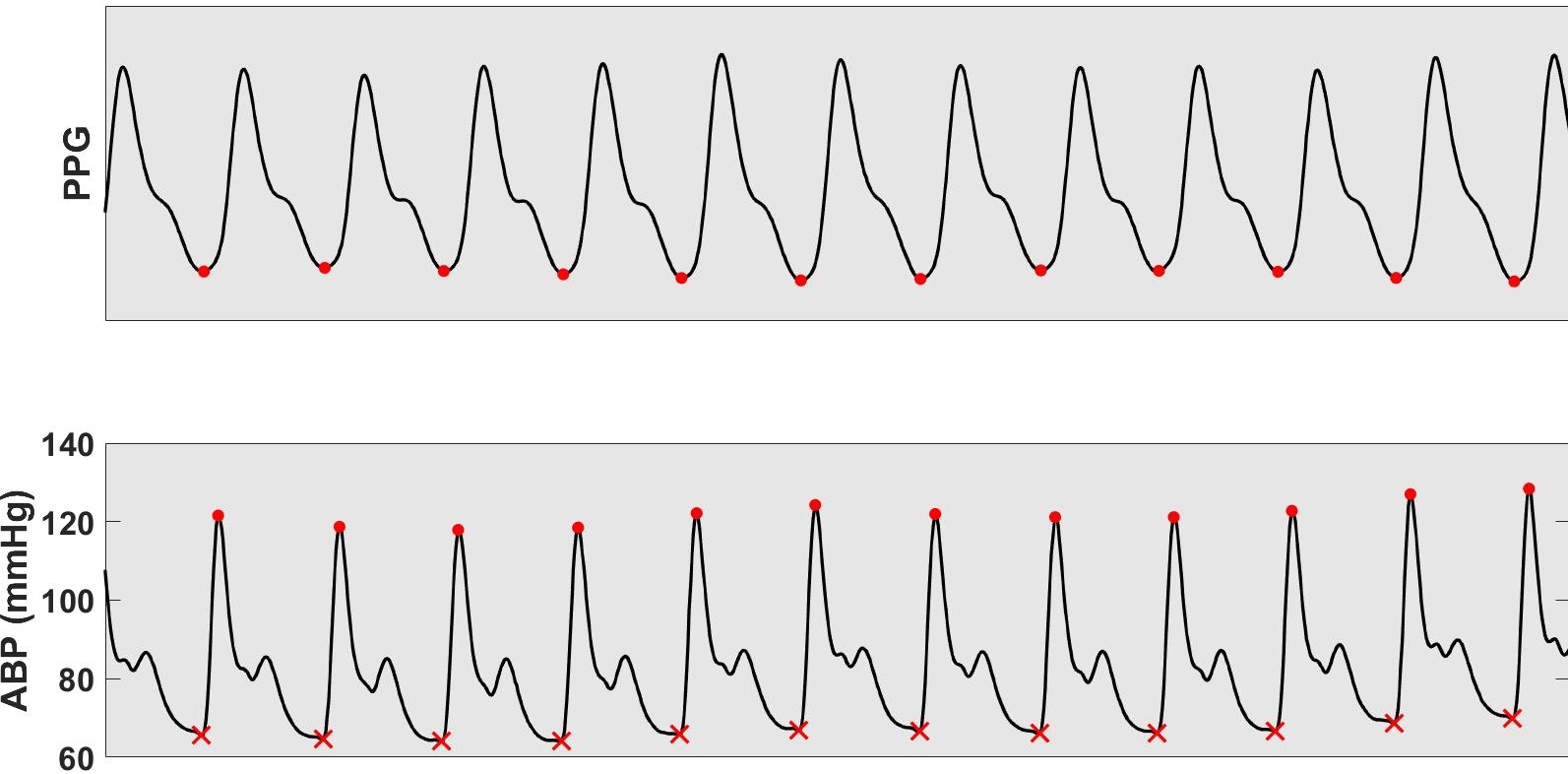}
\caption{PPG and ABP signals.}\label{Fig:Preprocessing}
\end{figure}
Before running the algorithm, the PPG signal should be segmented into PPG pulses, each representing one cardiac cycle. For this, the raw PPG signals\footnote{although the PPG signals in the MIMIC I records are already filtered, thus technically they are not considered raw PPG signals.} are filtered using a band-pass filter with cutoff frequencies of 0.3 Hz and 4.5 Hz. Then, the PPG pulses are identified using the beat-to-beat detection algorithm presented in \cite{Serj2020}. As illustrated in \cref{Fig:Preprocessing}, the troughs of the PPG signal are detected to segment it into PPG pulses. Each of the detected PPG pulses is further associated with a signal quality index (SQI) to indicate the quality of the detected pulse.  
\subsection{Reference Blood Pressure Data}\label{subsec:RefBP}
To generate the reference BP data, we detect the peaks of the ABP signal. The detected peak values correspond to the reference systolic blood pressure. Then, we identify the minimum values on the ABP waveform between the consecutive peaks. This minimum value corresponds to the reference diastolic blood pressure. Note that we visually inspect the ABP signal to discard noisy, abnormal, or clipped data and to ensure that the peaks and the troughs of the ABP are correctly detected. In \cref{Fig:Preprocessing}, we show an example of how the arterial SBP and DBP values are identified. For the arterial mean blood pressure (MBP), we use the following equation
\begin{equation}\label{eq:MBP}
\textbf{MBP} = \frac{1}{3}\textbf{SBP}+\frac{2}{3}\textbf{DBP}.
\end{equation}
The statistics of the arterial systolic and diastolic blood pressure values are summarized in \cref{Table:BPStats}.

\begin{table}
\caption{{The statistics of the arterial BP data.}}
\begin{center}
\scalebox{1}{\begin{tabular}{ l  c  c  }
\hline 
 & \textit{\textbf{Systolic BP}} & \textit{\textbf{Diastolic BP}}\\
\hline\vspace{0.1cm}
{\textbf{Mean (mmHg)}} & $121.47$ & $60.23$\\
%\hline
\vspace{0.1cm}{\textbf{SD (mmHg)}} & $21.48$ & $11.78$\\
%\hline
\vspace{0.1cm}{\textbf{Range (mmHg)}} & $[76.48, 199.12]$ & $[30.80, 105.53]$\\
\hline
\end{tabular}}\label{Table:BPStats}
\end{center}
\end{table}

\section{Algorithm}\label{sec:algorithm}
\begin{figure*}[h]
\centering
\includegraphics[width=0.85\textwidth]{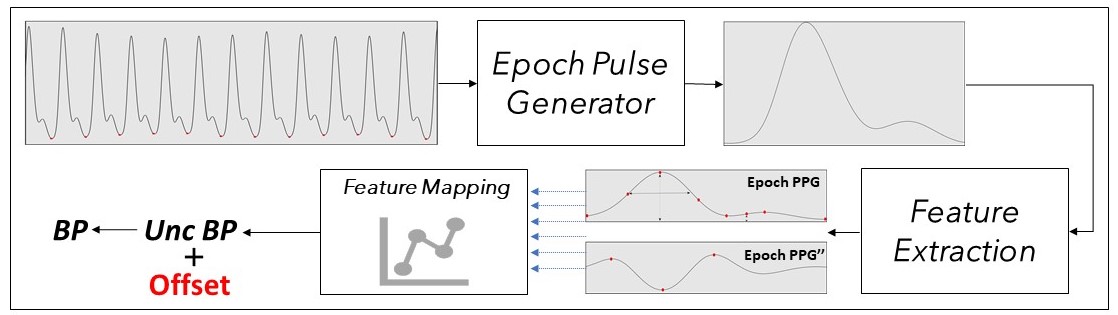}
\caption{The block diagram of the PPG-BPM algorithm: 1) epoch pulse generator producing an epoch pulse that represents the input PPG interval, 2) feature extraction block generating features from the epoch pulse, and 3) the feature mapping block outputting uncalibrated BP values. Using the zero-mean calibration process, an \textbf{offset} value is added to the uncalibrated BP values to generate the final BP estimates of the algorithm.}\label{Fig:AlgoBlocks}
\end{figure*}

%%%%%%%%%%%%%%%%%%%%%%%%%%%%%%%%%%%%%%%%%%%%%%%%%%%%%%%%%%%%%%%%%%%%
In this section, we describe our algorithm, which, as illustrated in \cref{Fig:AlgoBlocks}, consists of three main blocks: the epoch pulse generator block, the feature extraction block, and the feature mapping block.
\subsection{Epoch Pulse Generator}
The PPG signal is divided into 30-second intervals, called epochs. These epochs contain the PPG pulses identified during the signal pre-processing (refer to \cref{subsec:segmentation}). In a given epoch, if the number of high-quality PPG pulses (identified by the pulse SQI value) is above a certain threshold and the pulse width variation of these pulses is below another predefined threshold, the epoch is marked as \textit{clean}. Only \textit{clean} epochs are eligible to be processed by the epoch pulse generator block.

As the name suggests, the epoch pulse generator outputs one PPG pulse that accurately represents the input \textit{clean} epoch (see \cref{Fig:AlgoBlocks}). To achieve this goal, the given block normalizes the PPG pulses that have high SQI values to obtained PPG pulses with amplitude values of one. Each of the normalized PPG pulses are then re-sampled using 200 points. Finally, the re-sampled normalized PPG pulses are superimposed to generate the epoch PPG waveform.   
\subsection{Feature Extraction}
The feature extraction block extracts several time-related and amplitude-related features from the epoch PPG. In our algorithm, the feature extraction block extracts 27 features, out of which 16 features are extracted from the epoch PPG and 11 features from the second derivative (or the acceleration) of the epoch PPG (PPG" or APPG). The extracted features are based on 6 characteristic points identified on the PPG waveform (max. slope on the up-rise, the systolic peak, the half amplitude points, the dicrotic notch, the inflection point, and the diastolic peak) and 5 characteristic points identified on the PPG" waveform (waves $a$, $b$, $c$, $d$, and $e$ of the APPG) \cite{PWAFeaturesShimazaki2018}.
\subsection{Feature Mapping}
In this section, we explain how we construct the MLR models used to map the extracted features to SBP and DBP values. To generate the MLR coefficients, we exclusively use data from two records: records $041$ and $427$. The choice of these two records is based on three considerations: clean ABP and PPG data, sufficient variations in both the systolic and the diastolic blood pressure values, and a good generalization with respect to other data-sets\footnote{The MLR model generated using the two specified records generalizes well to other data-sets.}. The records $041$ and $427$ contribute with $1350$ and $1327$ \textit{clean} epochs, respectively. For each epoch, we compute reference arterial SBP and DBP values by averaging the beat-to-beat BP values generated (refer to \cref{subsec:RefBP}) within the interval of the given epoch. Moreover, each epoch is associated with 27 feature values extracted by the feature extraction block.

To generate the coefficients of the SBP MLR model, we use an iterative approach consisting of two iterations (can be extended to more iterations). During the first iteration, we use the data generated from the record $041$ to compute the MLR parameters that minimize the mean square error. As such, we get

\begin{equation}\label{eq:gamma1}
\overrightarrow{\gamma_{1}} = \arg\min_{\overrightarrow{\gamma}} \sum_{i=1}^n \left(\overrightarrow{\gamma}.\overrightarrow{f_i^{(1)}}-{y_i^{(1)}}\right)^2,
\end{equation}
where $\overrightarrow{\gamma_{1}}$ denotes the vector containing the MLR parameters, $n = 1350$ denotes the number of epochs in the record $041$, and $\overrightarrow{f_i^{(1)}}$ and ${y_i^{(1)}}$ denote the feature vector and the reference arterial SBP value of the $i$-th epoch of the given record, respectively.

\begin{figure}[h]
\centering
\includegraphics[width = \linewidth]{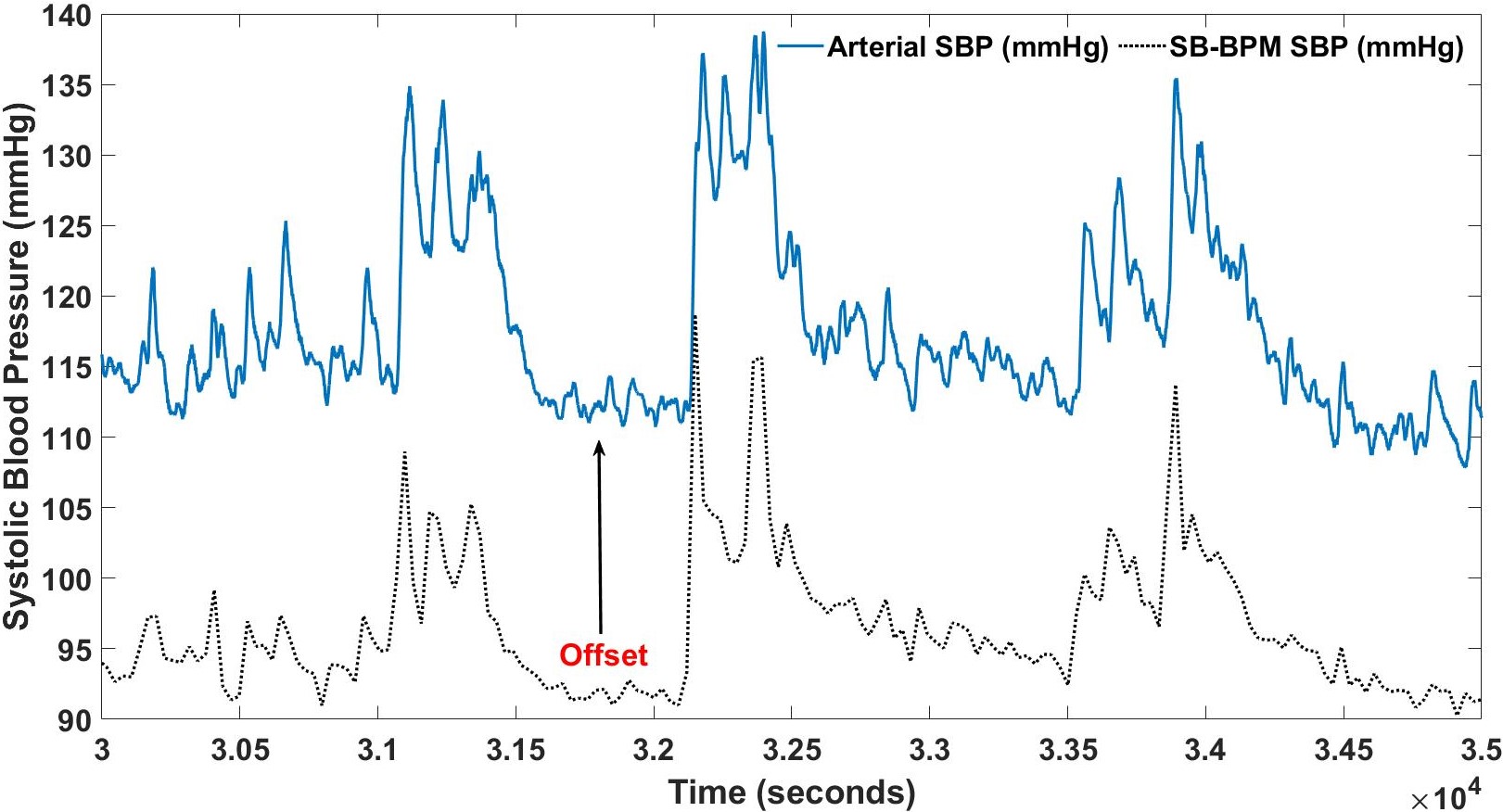}
\caption{Illustrating the fixed offset between the estimated SBP and the arterial SBP (Subject $427$).}\label{Fig:Calibration}
\end{figure}

After the first iteration, we use the generated MLR model (with the computed parameters $\overrightarrow{\gamma_{1}}$) on the epoch features of the record $427$. The estimated SBP values based on $\overrightarrow{\gamma_{1}}$ MLR model, as shown in \cref{Fig:Calibration}, accurately capture the variations of the arterial SBP with an offset  $\sim 20$ mmHg that remains stable throughout the recording. We compute this offset as follows 
\begin{equation}\label{eq:offset}
\textbf{offset} = \frac{1}{m} \sum_{i=1}^m \left(\overrightarrow{\gamma_1}.\overrightarrow{f_i^{(2)}}-{y_i^{(2)}}\right),
\end{equation}
where $m = 1327$ denotes the number of epochs in the record $427$, $\overrightarrow{f_i^{(2)}}$ denotes the feature vector of the $i$-th epoch of the record, and ${y_i^{(2)}}$ denotes the reference arterial SBP of the $i$-th epoch of the record. Note that in this work, we do not include information about the subjects (age, weight, body mass index, height, gender, etc.) as inputs to our algorithm. This means larger offsets are expected between the different data-sets. These offsets are compensated by the calibration process.  

For the second iteration, we want to use both records ($041$ and $427$) to regenerate the parameters of the MLR model. In \cref{eq:offset}, we computed the \textbf{offset} in the estimated SBP values of the record $427$ using the MLR parameters computed in \cref{eq:gamma1} based on the record $041$. For this reason, we compensate for this offset to compute the new set of MLR parameters as follows 

\begin{align}\label{eq:gamma2}\nonumber
\overrightarrow{\gamma_{2}} = \arg\min_{\overrightarrow{\gamma}} \Bigg(&\sum_{i=1}^n \left(\overrightarrow{\gamma}.\overrightarrow{f_i^{(1)}}-{y_i^{(1)}}\right)^2 + \\  
&\sum_{i=1}^m \left(\left(\overrightarrow{\gamma}.\overrightarrow{f_i^{(2)}}+\textbf{offset}\right)-{y_i^{(2)}}\right)^2\Bigg),
\end{align}
where $\overrightarrow{\gamma_{2}}$ is the vector containing the MLR parameters. These parameters are used by the  feature mapping SBP MLR model throughout all the data-sets analyzed in this work. The same approach is used to construct the DBP MLR model.

\section{Results}\label{sec:results}
%%%%%%%%%%%%%%%%%%%%%%%%%%%%%%%%%%%%%%%%%%%%%%%%%%%%%%%%%%%%%%%%%%%%
The metrics used to evaluate the accuracy of the proposed algorithm to estimate blood pressure are: mean absolute error (MAE, in mmHg), mean error (ME, in mmHg), standard deviation of error (SDE, in mmHg), percentage of estimated BP values with absolute error $\leq Y$ mmHg, $Y = (5, 10, 15)$, and the Pearson correlation coefficient $r$. 

On the other hand, the metrics used to evaluate the performance of the algorithm to classify BP values are:
\begin{align}\label{eq:Class}\nonumber
\textbf{Accuracy}(\%) &= 100\times\frac{\text{TP}+\text{TN}}{\text{TP}+\text{TN}+\text{FP}+\text{FN}}\\\nonumber
\textbf{Specificity}(\%) &= 100\times\frac{\text{TN}}{\text{TN}+\text{FP}}\\\nonumber
\textbf{Sensitivity}(\%) &= 100\times\frac{\text{TP}}{\text{TP}+\text{FN}}\\
\textbf{Precision}(\%) &= 100\times\frac{\text{TP}}{\text{TP}+\text{FP}},
\end{align}
where TP, TN, FP, and FN denote the number of `True Positives', `True Negatives', `False Positives', and `False Negatives,' respectively.

\subsection{Overall Performance Evaluation}
\begin{table}
\caption{{Overall performance evaluation over the $47153$ \textit{clean} 30-second epochs.}}
\begin{center}
\scalebox{1}{\begin{tabular}{ l  c  c  c }
\hline 
 & \textit{\textbf{Systolic BP}} & \textit{\textbf{Diastolic BP}} & \textit{\textbf{Mean BP}}\\
\hline\vspace{0.1cm}
{\textbf{MAE (mmHg)}} & $6.10$ & $4.65$ & $4.32$\\
%\hline
\vspace{0.1cm}{\textbf{ME (mmHg)}} & $0$ & $0$ & $0$\\
%\hline
\vspace{0.1cm}{\textbf{SDE (mmHg)}} & $8.08$ & $6.22$ & $5.75$\\
%\hline
\vspace{0.1cm}{\textbf{$\leq 5$ mmHg ($\% $)}} & $51.17$ & $63.65$ & $67.14$\\
%\hline
\vspace{0.1cm}{\textbf{$\leq 10$ mmHg ($\% $)}} & $82.61$ & $90.71$ & $91.74$\\
%\hline
\vspace{0.1cm}{\textbf{$\leq 15$ mmHg ($\% $)}} & $93.23$ & $97.03$ & $98.03$\\
%\hline
\vspace{0.1cm}{\textbf{Correlation r}} & $0.90$ & $0.85$ & $0.87$\\
\hline
\end{tabular}}\label{Table:MIMIC}
\end{center}
\end{table}
The overall performance of the PPG-BPM algorithm to estimate SBP, DBP, and MBP, is summarized in \cref{Table:MIMIC}. The obtained results confirm the good performance of the algorithm with SBP, DBP, and MBP MAE values of $6.10$ mmHg, $4.65$ mmHg, and $4.32$ mmHg, respectively, and SDE values of $8.08$ mmHg, $6.22$ mmHg, and $5.75$ mmHg, respectively. The ME values are equal to zero due to the zero-mean calibration process, explained in \cref{subsec:Calibration}. The results also show that more than $80\%$ of the SBP, the DBP, and the MBP estimates have an absolute error of less than 10 mmHg compared to the arterial BP values. 

\begin{figure*}[t]
\centering
\includegraphics[width = .8\textwidth]{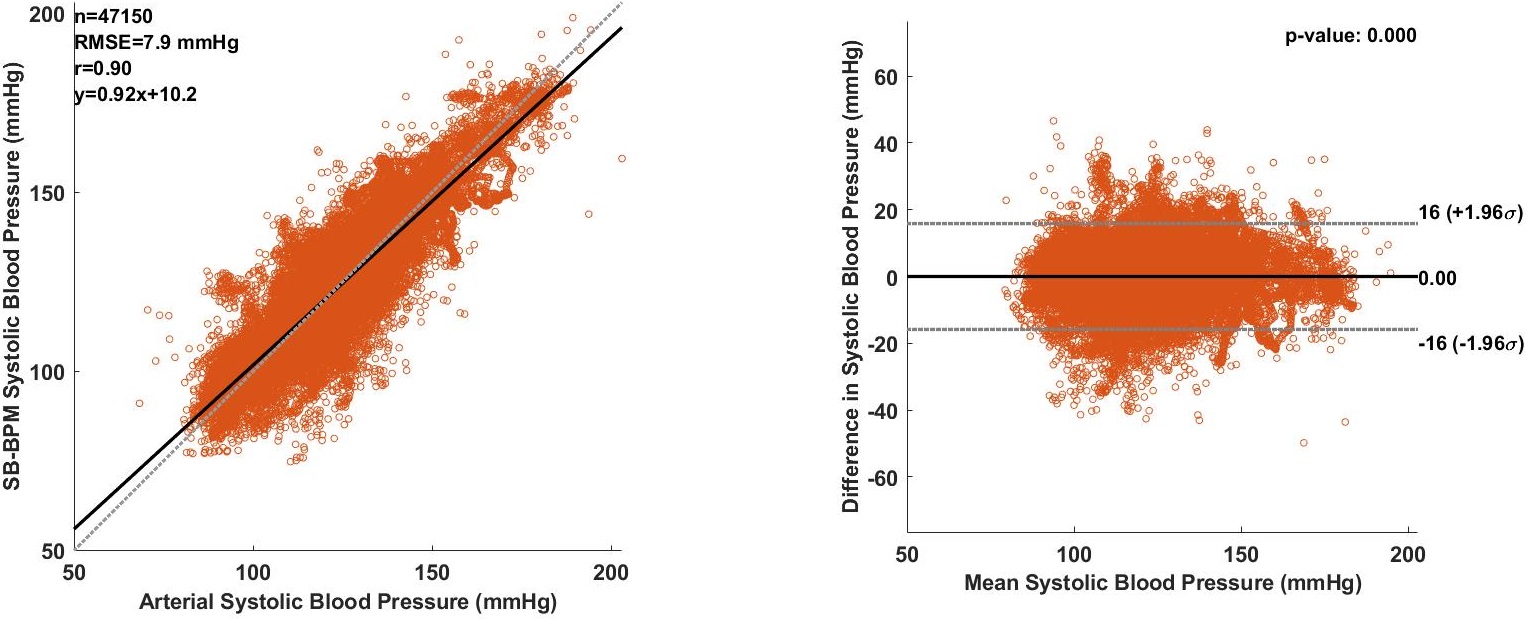}
\caption{The left plot shows the high correlation in the SBP values with Pearson correlation coefficient $r = 0.9$. The Bland-Altman plot to the right shows the $95\%$ confidence interval $\mu \pm 1.96\sigma = \lbrace-15.83$ mmHg, $15.83$ mmHg$\rbrace$.}\label{Fig:BA_SBP}
\end{figure*}
\begin{figure*}[t]
\centering
\includegraphics[width = .8\textwidth]{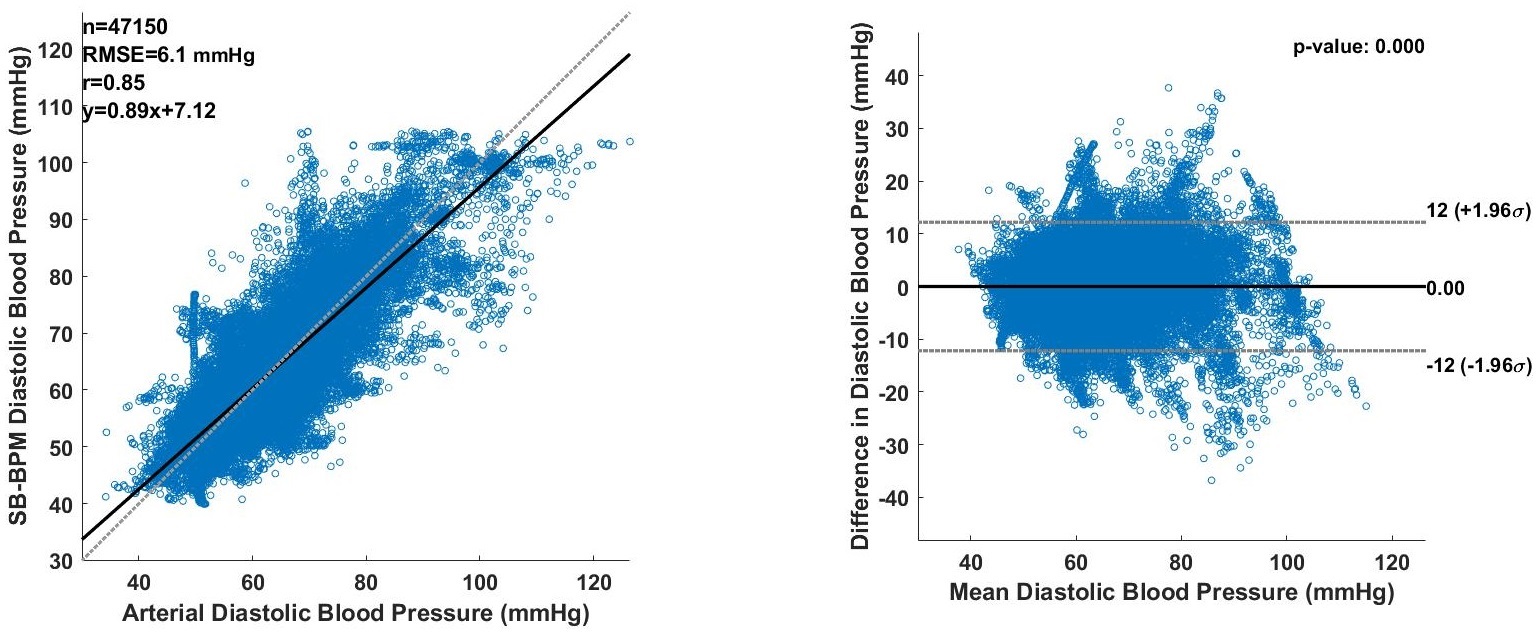}
\caption{The left plot shows the high correlation in the DBP values with Pearson correlation coefficient $r = 0.85$. The Bland-Altman plot to the right shows the $95\%$ confidence interval $\mu \pm 1.96\sigma = \lbrace-12.19$ mmHg, $12.19$ mmHg$\rbrace$.}\label{Fig:BA_DBP}
\end{figure*}

In \cref{Fig:BA_SBP} and \cref{Fig:BA_DBP}, we illustrate the high correlation between the PPG-based BP values generated by the proposed algorithm and the corresponding arterial BP values: $r$ values of $0.9$ and $0.85$, $p<.001$, for systolic and diastolic BP values, respectively. In the same figure, we show the $95\%$ limits of agreement in the Bland-Altman plots of the SBP and DBP values.

\subsection{Subject by Subject Performance Evaluation}
The PPG-BPM algorithm identifies the epochs that are clean and eligible for pulse wave analysis. The records with more than 15 minutes of \textit{clean} epochs ($>30$ \textit{clean} epochs) are included in the performance analysis. The number of epochs analyzed per subject and the corresponding statistics are summarized in \cref{Table:Subjects}. 
\begin{table}
\caption{Per subject performance evaluation.}
\begin{center}
\scalebox{1}{\begin{tabular}{ c c c c c c }
\hline 
 & & \multicolumn{2}{c}{\textit{\textbf{Systolic BP}}} & \multicolumn{2}{c}{\textit{\textbf{Diastolic BP}}} \\
 \hline 
 \textbf{Records} & \textbf{\# Epochs} & \textbf{MAE}& \textbf{SDE} & \textbf{MAE}& \textbf{SDE}\\
\hline
\vspace{0.1cm} {\textbf{039}} & $4154$ & $6.40$ & $8.36$ & $4.58$ & $5.58$ \\
\vspace{0.1cm} {\textbf{041}} & $1350$ & $6.22$ & $7.88$ & $5.52$ & $6.66$ \\
\vspace{0.1cm} {\textbf{055}} & $3083$ & $6.53$ & $8.38$ & $4.35$ & $6.09$ \\
\vspace{0.1cm} {\textbf{212}} & $2706$ & $7.27$ & $9.16$ & $4.41$ & $5.57$ \\
\vspace{0.1cm} {\textbf{224}} & $942$ & $6.55$ & $8.42$ & $6.54$ & $8.33$ \\
\vspace{0.1cm} {\textbf{225}} & $491$ & $4.80$ & $7.14$ & $6.01$ & $8.20$ \\
\vspace{0.1cm} {\textbf{226}} & $1109$ & $3.08$ & $3.75$ & $4.64$ & $6.34$ \\
\vspace{0.1cm} {\textbf{237}} & $1779$ & $6.04$ & $7.60$ & $4.84$ & $6.20$ \\
\vspace{0.1cm} {\textbf{240}} & $149$ & $5.14$ & $7.23$ & $10.22$ & $12.21$ \\
\vspace{0.1cm} {\textbf{252}} & $691$ & $9.15$ & $10.10$ & $8.89$ & $11.24$ \\
\vspace{0.1cm} {\textbf{253}} & $700$ & $6.27$ & $7.98$ & $4.66$ & $6.22$ \\
\vspace{0.1cm} {\textbf{254}} & $2907$ & $8.25$ & $9.30$ & $5.29$ & $6.43$ \\
\vspace{0.1cm} {\textbf{281}} & $965$ & $5.30$ & $6.78$ & $9.93$ & $10.17$ \\
\vspace{0.1cm} {\textbf{401}} & $50$ & $1.60$ & $2.21$ & $2.44$ & $3.29$ \\
\vspace{0.1cm} {\textbf{408}} & $570$ & $6.15$ & $8.43$ & $8.71$ & $9.66$ \\
\vspace{0.1cm} {\textbf{409}} & $248$ & $8.04$ & $9.17$ & $5.12$ & $6.53$ \\
\vspace{0.1cm} {\textbf{414}} & $446$ & $7.70$ & $9.09$ & $4.68$ & $5.56$ \\
\vspace{0.1cm} {\textbf{417}} & $117$ & $5.52$ & $6.57$ & $4.41$ & $5.70$ \\
\vspace{0.1cm} {\textbf{427}} & $1327$ & $4.70$ & $6.13$ & $3.21$ & $4.20$ \\
\vspace{0.1cm} {\textbf{430}} & $41$ & $9.77$ & $10.78$ & $2.76$ & $3.57$ \\
\vspace{0.1cm} {\textbf{439}} & $32$ & $6.69$ & $7.79$ & $3.94$ & $6.05$ \\
\vspace{0.1cm} {\textbf{443}} & $5377$ & $3.78$ & $4.47$ & $3.81$ & $5.21$ \\
\vspace{0.1cm} {\textbf{444}} & $2720$ & $7.17$ & $9.01$ & $4.27$ & $5.95$ \\
\vspace{0.1cm} {\textbf{449}} & $1905$ & $6.45$ & $8.53$ & $6.40$ & $8.60$ \\
\vspace{0.1cm} {\textbf{456}} & $2888$ & $3.82$ & $4.45$ & $5.67$ & $7.21$ \\
\vspace{0.1cm} {\textbf{466}} & $6987$ & $6.36$ & $8.11$ & $3.08$ & $3.89$ \\
\vspace{0.1cm} {\textbf{474}} & $2162$ & $7.16$ & $8.91$ & $2.99$ & $3.75$ \\
\vspace{0.1cm} {\textbf{482}} & $1257$ & $7.98$ & $9.01$ & $5.14$ & $6.12$ \\
\hline
\end{tabular}}\label{Table:Subjects}
\end{center}
\end{table}
The obtained results confirm that the algorithm maintains a good performance over the majority of the records. For the SBP, the MAE and the SDE values are $\sim 6$ mmHg and $\sim 8$ mmHg, respectively, for the majority of the records. For the DBP, the MAE and the SDE values are $\sim 4.5$ mmHg and $\sim 6$ mmHg, respectively, for the majority of the records. These values reflect the overall performance of the proposed algorithm shown in \cref{Table:MIMIC}.

\subsection{Classification Results}
\label{subsec:ClassResults}
One of the main goals of continuous blood pressure monitoring is to detect elevated blood pressure values at early stages to prevent future complications. For this reason, it is relevant to study the accuracy of PPG-based methods in identifying high BP values. 

In this study, we classify the BP measurements into two groups: \textit{hypertensive} and \textit{non-hypertensive}. Our goal is to detect hypertension or the risk of developing it. The Stage I Hypertension (pre-hypertension) is characterized by $130\leq \text{SBP} < 140$ mmHg and/or $80\leq \text{DBP} < 90$ mmHg. For this reason, we classify the BP values as \textit{hypertensive} if SBP $\geq 130$ mmHg and/or DBP $\geq 80$ mmHg, and as \textit{non-hypertensive}, otherwise.

\begin{figure}[h]
\centering
\includegraphics[width = 0.8\linewidth]{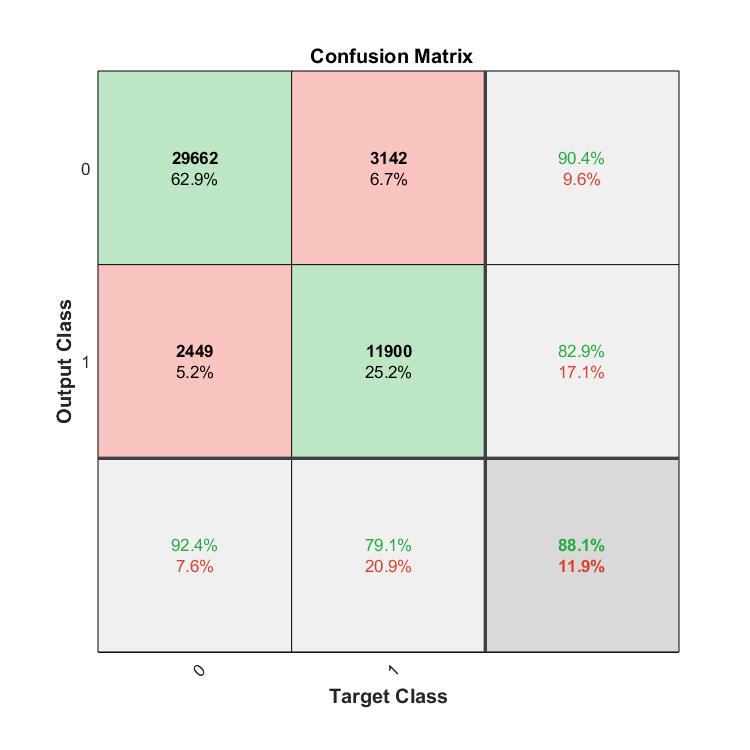}
\caption{True and false classifications identified by green and red blocks, respectively. }\label{Fig:ConfusionMatrix}
\end{figure}

The target classes of the $47153$ epochs are derived using the arterial BP measurements and the output classes are derived from the PPG-based BP estimates of our algorithm. Based on the confusion matrix shown in \cref{Fig:ConfusionMatrix}, $\text{TP} = 11900$, $\text{TN} = 29662$, $\text{FP} = 2449$, and $\text{FN} = 3142$. Using \cref{eq:Class}, we obtain the classification results shown in \cref{Table:Accuracy}. The results show the good performance of the PPG-based classification.

\begin{table}
\caption{Performance metrics of the classification of the $47153$ \textit{clean} epochs.}
\begin{center}
{\begin{tabular}{  l c c  }
{\textit{Classification Metrics}} & & $\% $\\
\hline 
{\textbf{Accuracy }} & & $88.14$ \\
%\hline
{\textbf{Specificity}} & & $92.37$\\
%\hline
{\textbf{Sensitivity}} & & $79.11$ \\
%\hline
{\textbf{Precision}} & & $82.93$ \\
\hline
\end{tabular}}\label{Table:Accuracy}
\end{center}
\end{table}

We further examine the accuracy of our algorithm within each of the two specified groups: \textit{hyp.} group ($15042$ epochs)  and \textit{non-hyp.} group ($32111$ epochs). The results depicted in \cref{Table:Classification} show that the algorithm achieves a good performance for both the \textit{hyp.} and the \textit{non-hyp.} groups with a slightly better performance for the latter. This is due to the fact that high blood pressure deteriorates the PPG morphology and renders the feature extraction more challenging.

\begin{table}
\caption{{Performance Evaluation over different BP groups.}}
\begin{center}
\scalebox{1}{\begin{tabular}{ l  c c c  c  }
\hline 
 & \multicolumn{2}{c}{\textit{\textbf{Systolic BP}}} & \multicolumn{2}{c}{\textit{\textbf{Diastolic BP}}} \\
 \hline 
 & \textit{\textbf{Hyp.}}& \textit{\textbf{non Hyp.}} & \textit{\textbf{Hyp.}}& \textit{\textbf{non Hyp.}}\\
\hline\vspace{0.1cm}
{\textbf{MAE (mmHg)}} & $6.76$ & $5.78$ & $5.32$ & $4.33$ \\
%\hline
\vspace{0.1cm}{\textbf{ME (mmHg)}} & $-2.23$ & $1.04$ & $-1.74$ & $0.81$ \\
%\hline
\vspace{0.1cm}{\textbf{SDE (mmHg)}} & $8.51$ & $7.65$ & $6.92$ & $5.68$\\
%\hline
\vspace{0.1cm}{\textbf{$\leq 5$ mmHg ($\% $)}} & $47.35$ & $52.96$ & $57.63$ & $66.48$ \\
%\hline
\vspace{0.1cm}{\textbf{$\leq 10$ mmHg ($\% $)}} & $76.65$ & $85.40$ & $85.79$ & $93.01$ \\
%\hline
\vspace{0.1cm}{\textbf{$\leq 15$ mmHg ($\% $)}} & $90.53$ & $94.49$ & $95.06$ & $97.96$ \\
\hline
\end{tabular}}\label{Table:Classification}
\end{center}
\end{table}

\section{Discussions}\label{sec:discussions}
%%%%%%%%%%%%%%%%%%%%%%%%%%%%%%%%%%%%%%%%%%%%%%%%%%%%%%%%%%%%%%%%%%%%
\subsection{Performance with respect to Cuff measurements}
It is important to compare the performance of our algorithm with respect to cuff-based recordings, which is the gold-standard for non-invasive BP measuring. For this purpose, we utilize the University of Queensland Vital Signs data-set \cite{QUD}.
\subsubsection{Queensland University Database (QUD)}
\begin{figure} 
\centering
\includegraphics[width = \linewidth]{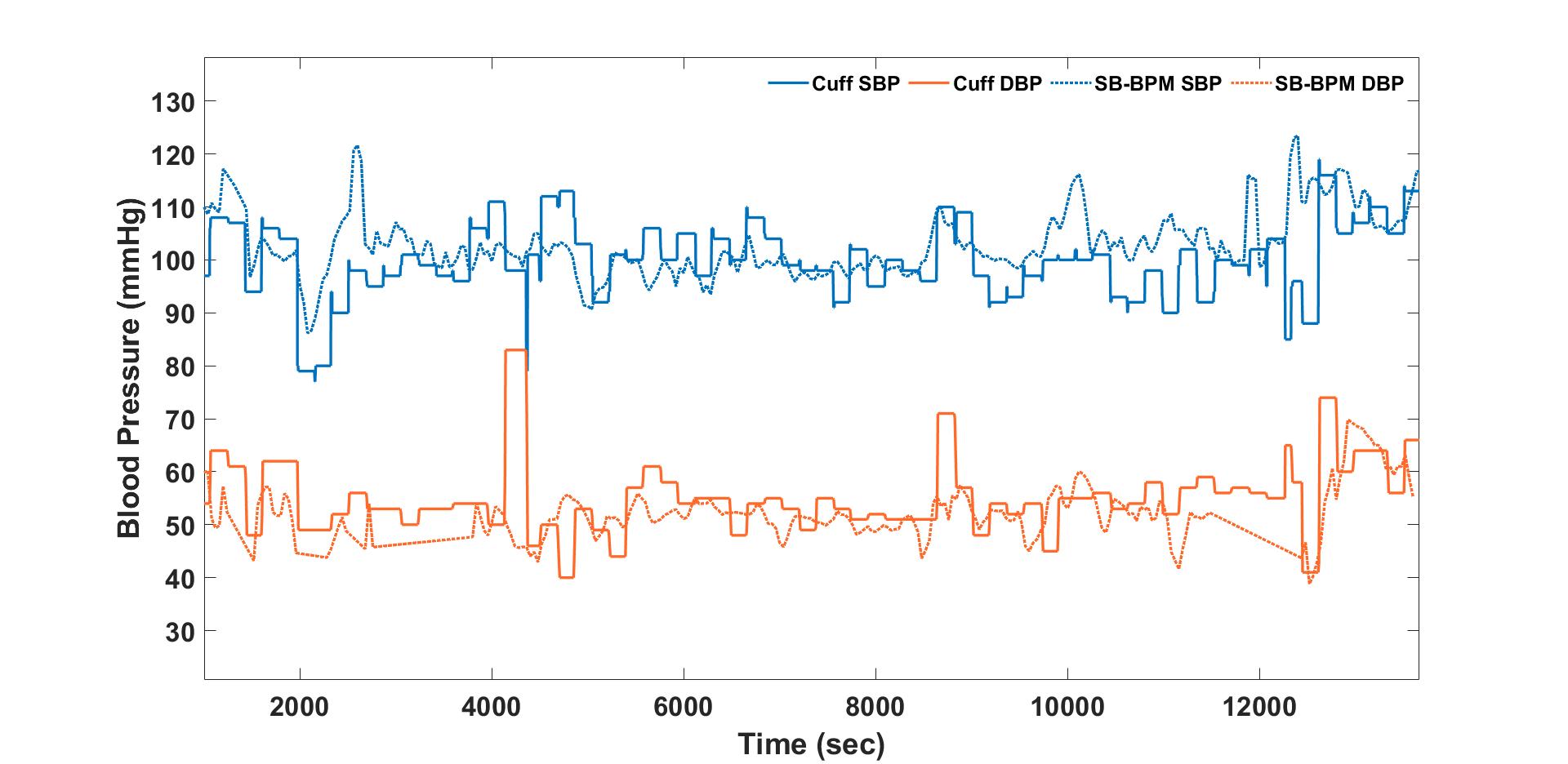}
\caption{SBP and DBP plots for both the cuff-based measured values and PPG-based PPG-BPM estimates (QUD case $03$).}\label{Fig:QUD}
\end{figure}

The PPG signal from the $32$ available data-sets is pre-processed in a similar fashion to what we did for the MIMIC I database. The algorithm is ran on the processed PPG data to generate the \textit{clean} 30-second epochs.
To evaluate the accuracy of the proposed algorithm, we consider the \textit{clean} epochs that contain cuff recordings. This resulted in $1709$ epochs used for the performance evaluation. In \cref{Fig:QUD}, we illustrate the PPG-based SBP and DBP estimation of our algorithm together with the QUD cuff-based recordings. The plot shows how close the BP estimates from the algorithm are to the cuff measurements. It is important to recall that the same MLR parameters ($\overrightarrow{\gamma_2}$) generated in \cref{eq:gamma2} are used for the QUD data-sets.

\begin{table}
\caption{{Overall Performance over the 1709 \textit{clean} 30-second epochs}}
\begin{center}
\scalebox{1}{\begin{tabular}{ l  c  c  }
\hline 
 & \textit{\textbf{Systolic BP}} & \textit{\textbf{Diastolic BP}}\\
\hline\vspace{0.1cm}
{\textbf{MAE (mmHg)}} & $6.70$ & $6.64$\\
%\hline
\vspace{0.1cm}{\textbf{ME (mmHg)}} & $1.50$ & $0.40$\\
%\hline
\vspace{0.1cm}{\textbf{SDE (mmHg)}} & $8.80$ & $8.00$\\
%\hline
\vspace{0.1cm}{\textbf{$\leq 5$ mmHg ($\% $)}} & $49.30$ & $50.67$\\
%\hline
\vspace{0.1cm}{\textbf{$\leq 10$ mmHg ($\% $)}} & $76.75$ & $80.45$\\
%\hline
\vspace{0.1cm}{\textbf{$\leq 15$ mmHg ($\% $)}} & $90.58$ & $89.46$\\
\hline
\end{tabular}}\label{Table:QUD}
\end{center}
\end{table}
\subsubsection{Results}
In \cref{Table:QUD}, we show that the PPG-BPM achieves a good performance compared to cuff-based BP measurements. The proposed algorithm estimates the systolic and the diastolic blood pressures with a mean error of $1.50\pm8.80$ mmHg and $0.40 \pm 8.00$ mmHg, respectively. The results also show that more than $76\%$ of both the SBP and the DBP estimates have an absolute error of less than $10$ mmHg compared to the cuff-based BP recordings. Note that for this study, we calibrate the output of the algorithm based on the first cuff measurement, which results in non-zero ME values.

For both the SBP and the DBP estimates, the accuracy metrics (ME $\pm$ SDE) of the PPG-BPM are very close to the ISO/ANSI/AAMI protocol $5 \pm 8$ mmHg requirement.

\subsection{HR vs BP}
\begin{figure} 
\centering
\includegraphics[width = \linewidth]{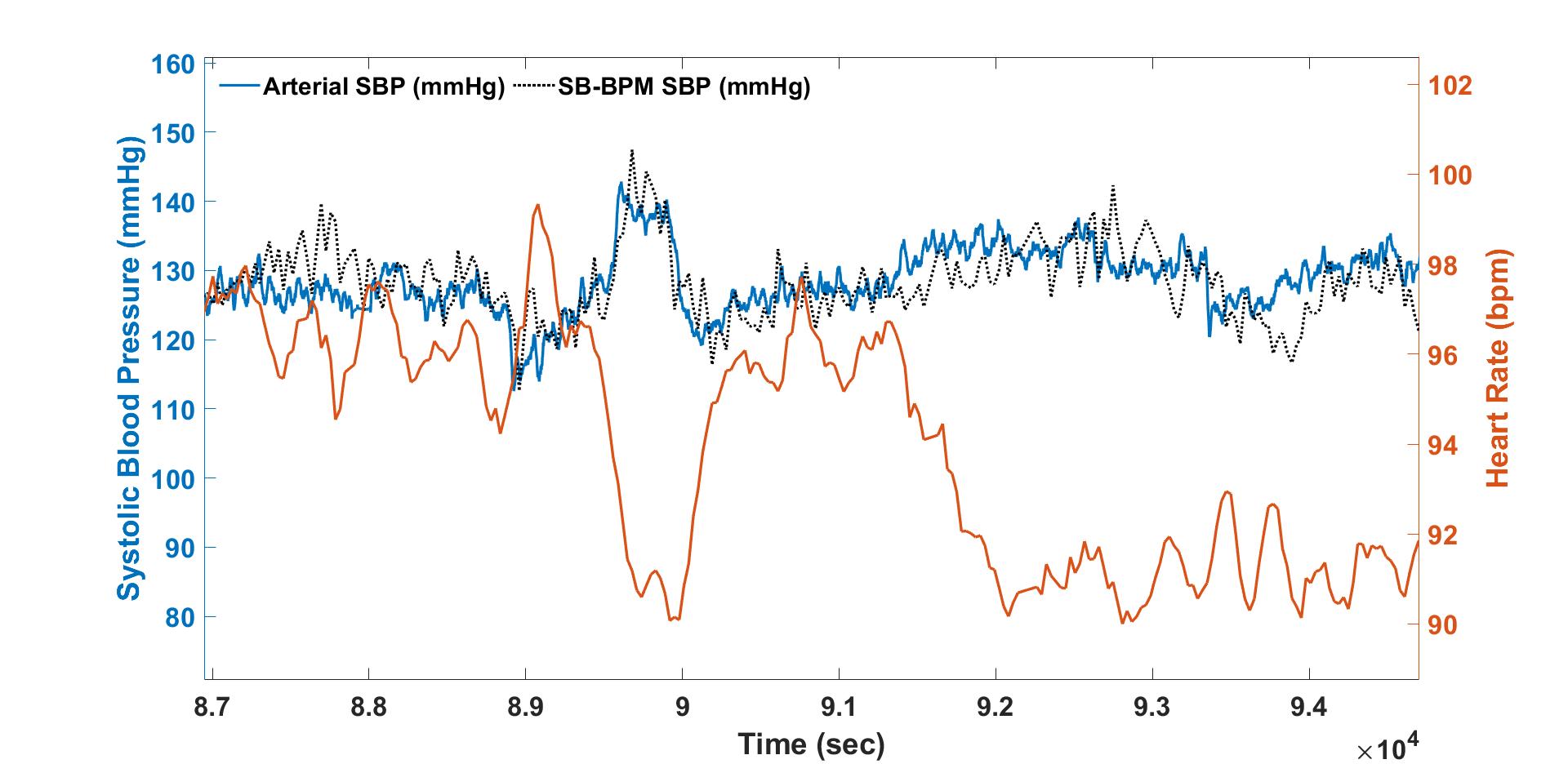}
\caption{Systolic BP variations vs HR variations (MIMIC I record $254$).}\label{Fig:SBP_vs_HR}
\end{figure}
One of the methods to obtain continuous BP values using intermittent cuff measurements is to interpolate the BP values using the heart rate variations. This can be continuously and accurately monitored using the subject's ECG or PPG. Despite the fact that the heart rate often varies in agreement with the blood pressure, it cannot act as a reliable substitute. In \cref{Fig:SBP_vs_HR}, we plot a small interval (taken from the MIMIC I record $254$) of the HR variations and the SBP variations. The plot illustrates how the systolic blood pressure varies independently and sometimes quite reversely to the heart rate. Moreover, the plot shows how our algorithm is capable of accurately estimating the arterial SBP values. The objective of this discussion is to reveal how the algorithm is not biased by the heart rate variations.

\subsection{Calibration}\label{subsec:Calibration}
The SBP and the DBP estimates from the algorithm are un-calibrated (see \cref{Fig:AlgoBlocks}). For the MIMIC I data-sets, to calibrate the BP estimates, we adopt the zero-mean calibration. This approach introduces SBP and DBP offsets so that the mean errors (ME) with respect to the arterial SBP and DBP values are equal to zero. On the other hand, for the QUD data-sets, we adopt the starting-point calibration. This approach introduces the SBP and DBP offsets based on the first cuff-based SBP and DBP measurements. Note that in both the zero-mean calibration and the starting-point calibration, a single SBP offset and a single DBP offset are used per record/subject. 

\section{Conclusion}\label{sec:conclusion}
In this work, we evaluate the performance of the PPG-BPM algorithm for continuous blood pressure monitoring on the MIMIC I data-sets. The obtained results over a total of $47153$ 30-second epochs ($\sim 393$ hours of data) confirm that the proposed algorithm is capable of accurately estimating systolic and diastolic blood pressure. For the SBP, the algorithm achieves a mean absolute error of $6.10$ mmHg and an error standard deviation of $8.08$ mmHg compared to the arterial line BP values. For the DBP, the algorithm achieves a mean absolute error of $4.65$ mmHg and an error standard deviation of $6.22$ mmHg. The proposed algorithm with its good performance answers the increasing demand for accurate non-invasive continuous BP monitoring systems, which helps reduce the prevalence of hypertension worldwide.
%%%%%%%%%%%%%%%%%%%%%%%%%%%%%%%%%%%%%%%%%%%%%%%%%%%%%%%%%%%%%%%%%%%%

\bibliography{bibfile}
\bibliographystyle{ieeetr}
\end{document}